\documentclass[onefignum,onetabnum]{siamart171218}

\usepackage{lipsum}
\usepackage{amsfonts}
\usepackage{graphicx}
\usepackage{epstopdf}
\usepackage{algorithmic}
\ifpdf
  \DeclareGraphicsExtensions{.eps,.pdf,.png,.jpg}
\else
  \DeclareGraphicsExtensions{.eps}
\fi

\newsiamremark{remark}{Remark}
\newsiamremark{hypothesis}{Hypothesis}
\crefname{hypothesis}{Hypothesis}{Hypotheses}
\newsiamthm{claim}{Claim}

\headers{Discrete Line Fields on Surfaces}{T. Novello, J. Paix\~{a}o, C. Tomei, T. Lewiner}

\title{Discrete Line Fields on Surfaces}

\author{Tiago Novello \thanks{PUC-Rio and IMPA, Rio de Janeiro
  (\email{tiago.novello@impa.br}).}
\and Jo\~{a}o Paix\~{a}o\thanks{UFRJ/DCC, Rio de Janeiro
  (\email{jpaixao@dcc.ufrj.br}).}
\and Carlos Tomei\thanks{PUC-Rio, Rio de Janeiro  
  (\email{tomei@mat.puc-rio.br}).}
  \and Thomas Lewiner\thanks{Ceva Santé Animale, Paris
  (\email{lewiner@gmail.com}).}}

\usepackage{amsopn}

\makeatletter
\newcommand*{\addFileDependency}[1]{
  \typeout{(#1)}
  \@addtofilelist{#1}
  \IfFileExists{#1}{}{\typeout{No file #1.}}
}
\makeatother

\newcommand*{\myexternaldocument}[1]{%
    \externaldocument{#1}%
    \addFileDependency{#1.tex}%
    \addFileDependency{#1.aux}%
}

\ifpdf
\hypersetup{
  pdftitle={Discrete Line Fields on Surfaces},
  pdfauthor={T. Novello, J. Paix\~{a}o, C. Tomei, T. Lewiner}
}
\fi

\newcommand{\K}{\mathcal{S}}
\newcommand{\M}{\mathcal{X}}

\renewcommand{\L}{\mathcal{L}}

\myexternaldocument{ex_supplement}

\begin{document}

\maketitle

\begin{abstract}
Vector fields and line fields, their counterparts without orientations on  tangent lines, are familiar objects in the theory of dynamical systems. Among the techniques used in their study, the Morse--Smale decomposition of a (generic) field plays a fundamental role, relating the  geometric structure of phase space to a combinatorial object consisting of critical points and separatrices. Such concepts led Forman to a satisfactory theory of discrete vector fields, in close analogy to the continuous case. 

In this paper, we introduce discrete line fields. Again, our definition is rich enough to provide the counterparts of the basic results in the theory of continuous line fields: an Euler-Poincar\'e formula, a Morse--Smale  decomposition and a topologically consistent cancellation of critical elements, which allows for topological simplification of the original discrete line field.
\end{abstract}




\vspace{1cm}
\section{Introduction} \label{introduction}\hfill\\
Since Poincar\'e, the study of vector fields on manifolds make use of geometric and topological methods. Morse--Smale decompositions of phase space, consisting of special objects which include critical points and separatrices, are fundamental in a number of theoretical and applied approaches. 

Inspired by the concepts associated with the Morse--Smale structure, Forman introduced discrete vector fields with several combinatorial properties, whose theory runs pretty much in parallel with the theory for the continuous case. The fertility of these ideas is clear from the many resulting applications in recent years.

In another more traditional extension, the continuous theory has been extended to encompass line fields, which also have shown a strong potential for applications. 
In this paper, we introduce discrete line fields. As we shall see, discrete vector fields are special cases of this new definition, as (continuous)  vector fields are special cases of line fields. As in Forman's theory, our definition is rich enough to provide the counterparts to the basic results of the previous contexts: an Euler--Poincar\'e formula (Theorem \ref{theorem:euler_formula}), the existence of a Morse--Smale decomposition of the line field (Theorem \ref{theorem:MS_decomposition_line}) and techniques of cancellation of critical elements similar to Forman's \cite{forman98_1}, widely used in applications \cite{huang2008spectral,jan11,weinkauf2010topology, weinkauf2005}. 

Once the underlying objects are appropriately defined, our main result is Theorem \ref{theorem:homotopy}, a line field version of Forman's homotopy theorem:  every discrete line field corresponds to a  simpler line field consisting only of critical cells. 

The discrete formulation may be helpful in overcoming numerical issues raised in the application of line fields \cite{sreevalsan2011eigenvector, tricoche2002,tricoche2001_1,tricoche2001,wang2017robustness} which are not properly tractable using vector field. The subject offers a wealth of opportunities in combinatorial analysis and requires efficient algorithms dealing with the topology of discrete line fields.

\newpage
\section{Continuous Line Fields}\hfill\\
A \textit{line field} on a surface is a smooth map which assigns a tangent line to all but a finite number of points. Thus, for example, the two eigenlines associated to a generic 2D symmetric tensor field generate two such  fields. Line fields model a number of physical properties, like velocity and temperature gradient in fluid flow \cite{helman1991visualizing}, stress, and momentum flux in elasticity \cite{delmarcelle1993visualizing}. In~computer graphics and visualization, line fields are often studied in topological segmentation of curvature fields \cite{dong2006, huang2008spectral, kalberer2007quadcover}, remeshing \cite{alliez2003, marinov2006robust, renata15}, and come up in the visualization of vector/line/symmetric tensor fields \cite{sreevalsan2011eigenvector,tricoche2002, wang2017robustness, auer2011complete,delmarcelle94, knoppel2015stripe, nascimento2010topology}. 

As for vector fields, line fields give rise to \textit{orbits}  ~\cite{tricoche2002,delmarcelle94}.
We take a topological approach, grouping orbits with similar behavior \cite{weinkauf2010topology,nikolaev97,cazals2003molecular, peixoto73,robins2011theory,weinkauf2008}. The theory usually considers generic fields, the so called \textit{Morse--Smale  fields} \cite{peixoto73, andronov1937,nikolaev2013foliations}, whose stability properties are especially benign from the numerical point of view \cite{tricoche2002, delmarcelle94}. For such fields, there is a \textit{Morse--Smale decomposition}, consisting of vertices given by critical points (points in the surface with no tangent line attached to them) and separatrices, which are orbits connecting critical points.

\begin{figure}[ht]
	\begin{center}
			\begin{tabular}{cc}
				\includegraphics[width=3.5cm]{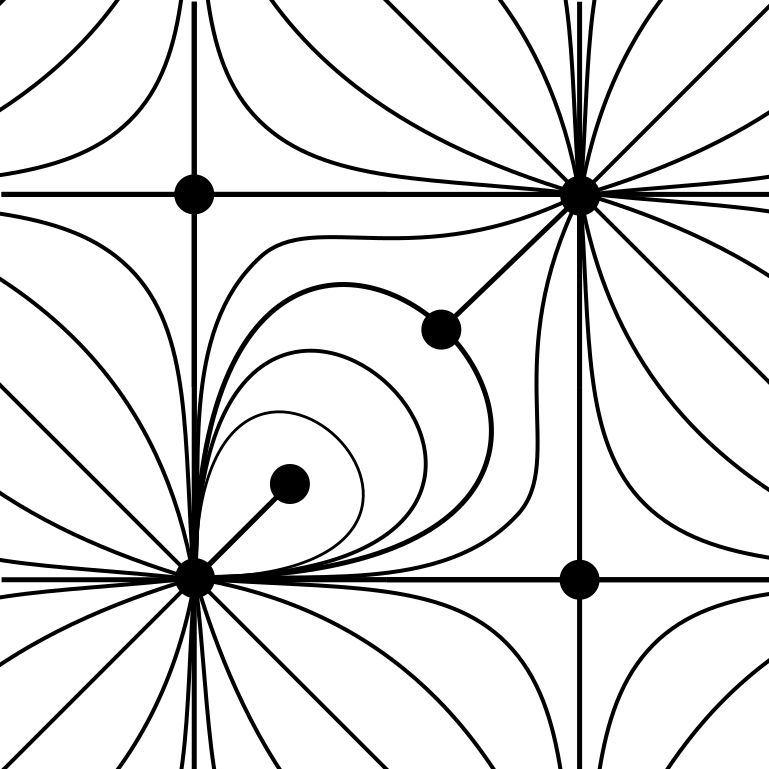}&
				\includegraphics[width=3.5cm]{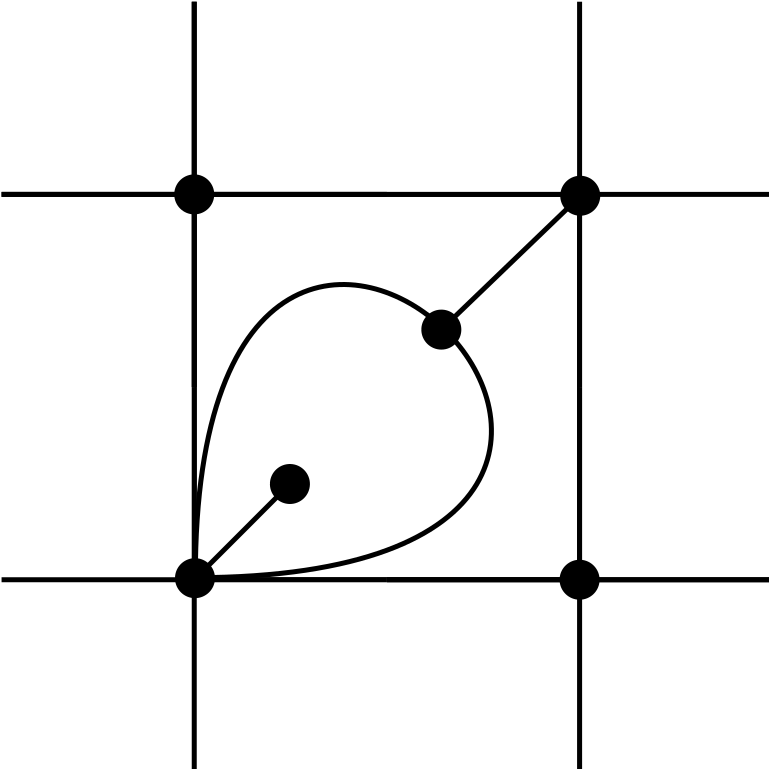}\\
				(a)&(b)
		\end{tabular}
		\caption{(a) A line field: dots are  critical points, (b) Critical points and  separatrices decompose the phase space in regions. }
		\label{figure:high_func}
	\end{center}
\end{figure}

Line fields are more general than vector fields: in Fig.~\ref{figure:high_func}(a), the field in the neighborhood of the two critical elements, in the center, can not be oriented into a vector field. These critical points are non-orientable and admit no counterpart in Forman's definition of a discrete vector field~\cite{forman98}.~However, these critical points admit discrete correspondents in our line field definition.

\vspace{0.5cm}
\section{Discrete vector fields}\label{discretevectorfields}\hfill\\
We recall the basic facts of Forman's theory of discrete vector fields. We follow the notation in \cite{gross1987topological}. An \textit{embedding} $i:G \to S$ of a graph $G$ into a connected, compact surface $S$ is a $1$-$1$ continuous map. 
A {\it 2-cell embedding} $i: G \to S$ is an embedding for which the components of the set $S - i(G)$ are homeomorphic to open discs, the \textit{faces}. 
A 2-cell embedding $i:G \to S$ induces a {\it CW decomposition} $\K=(V,E,F)$ of $S$, where $i(G)=(V,E)$ in the obvious way and $F$ are the faces of $S- i(G)$. 
The \textit{dimension} of a cell $\sigma$ in $\K$, denoted by $\dim(\sigma)$, is $0$, $1$ or $2$, if $\sigma$ is a vertex, edge or face, respectively. Vertices, edges, and faces in a CW decomposition are also denoted by $0$-cells, $1$-cells, and $2$-cells.
Given a face $\Sigma$, the edges  in its boundary  can be arranged along a closed loop  $\sigma_1,\sigma_2,\ldots,\sigma_k$, forming the \textit{boundary walk} of $\Sigma$. 

\begin{figure}[ht]
\begin{center}
\begin{tabular}{ccc}
	\includegraphics[width=2.4cm]{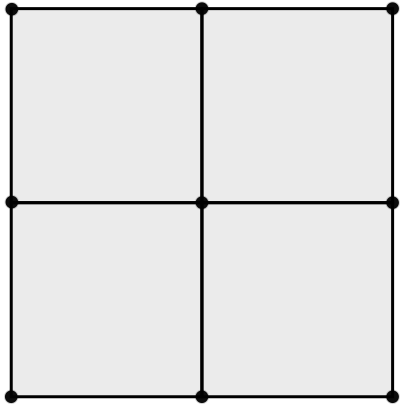}&
	\includegraphics[width=2.4cm]{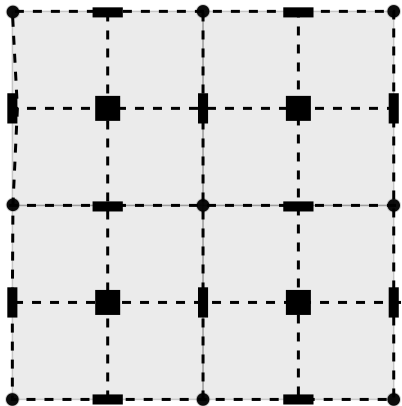}&
	\includegraphics[width=2.4cm]{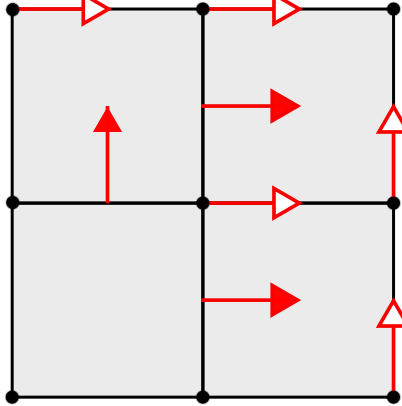}\\
	(a)&(b)&(c)
\end{tabular}\vspace{-0.1cm}
	\caption{(a) A CW decomposition of a neighborhood of a surface. (b) The Hasse diagram: notice the cells in three different dimensions. (c) A discrete vector field.}
	\label{figure:MorseMatching}
\end{center}
\end{figure}

The {\it Hasse diagram} $H(\K)$ of $\K$ is a graph embedding whose vertices  consist of a unique point in each cell of $\K$, and whose edges are  disjoint lines  connecting  points of adjacent cells of dimension differing by one (Fig. \ref{figure:MorseMatching}(a) and (b)). We abuse language slightly and think of the vertices of $H(\K)$ as the cells of $\K$. 

The \textit{dual} CW decomposition $\K^*=(V^*,E^*,F^*)$ of $\K$ is another CW decomposition of $S$. Its set of vertices $V^*$ consists of a vertex in the interior of each face of $\K$. We create an edge in $E^*$ for each pair of adjacent faces in $K$. The faces $F^*$ is the set $S-(V^*, E^*)$, which correspond to the set of vertices in $\K$. By construction, $H(\K^*)=H(\K)$ .

A {\it matching in $H(\K)$} is a collection $\M$ of disjoint edges in the Hasse diagram $H(\K)$. Forman defines a {\it discrete vector field} as a pair $(\K,\M)$ \cite{forman98}.  We interpret an element $\{\sigma,\tau\} \in \M$ with $dim(\sigma)+1=dim(\tau)$ as an arrow running from $\sigma$ to $\tau$, like the red arrows in Fig.~\ref{figure:MorseMatching}(c).  
As $\K$ and $\K^*$ have mirrored Hasse diagrams, we define the \textit{dual} discrete vector field $(\K^*,\M^*)$ as the mirrored matching $\M^*$ induced by $\M$. A matched pair $\{\sigma,\tau\}\in\M$, with $dim(\sigma)+1=dim(\tau)$ gives rise to a pair $\{\tau^*, \sigma^*\}\in\M^*$ such that $dim(\sigma^*)=dim(\tau^*)+1$: dualization inverts the direction of the arrows, the discrete counterpart of reversing vector orientation in a continuous vector field.

\begin{figure}[ht]
\begin{center}
\begin{tabular}{ccc}
	\raisebox{.24\height}{\includegraphics[width=1.7cm]{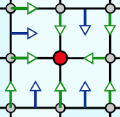}}&
	\includegraphics[width=1.7cm]{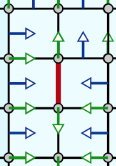}&
	\includegraphics[width=2.4cm]{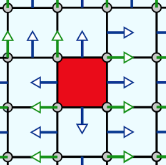}\\
	(a)&(b)&(c)
\end{tabular}\vspace{-0.1cm}
	\caption{(a) A critical vertex. (b) A critical edge. (c) A critical face.}
	\label{figure:criticals_vector}
\end{center}
\end{figure}

An unmatched cell in a discrete vector field is a \textit{critical cell} and its \textit{index}, denoted by $index(\sigma)$, is $(-1)^{dim(\sigma)}$ for a critical cell and $0$ otherwise (Fig. \ref{figure:criticals_vector}).

Each matched pair $\{\sigma, \tau\}$ in a discrete line field $(\K,\M)$ contributes with zero to the Euler characteristic $\chi(S)=|V|-|E|+|F|$ of the  surface $S$, as their dimensions differ by one. As the remaining cells are critical, a discrete version of the Poincar\'e-Euler formula \cite{forman98} holds.

\begin{theorem}[Euler-Poincar\'e]
\label{theorem:forman_euler_formula} 
Let $(\K,\M)$ be a discrete vector field of a surface $S$. Then
 $$\chi(S)=\displaystyle \sum_{\sigma\in \K}index(\sigma)\ .$$
\end{theorem}

The basic ingredient in the construction of a \textit{Morse--Smale decomposition} of $(\K,\M)$ is Forman's definition 
\cite{forman98_1} of a \textit{$\M$-path} of dimension $p$,  a sequence  of $p$-cells in $\K$,
$\gamma=\sigma_1\sigma_2\sigma_3\ldots\sigma_k,$
such that for each $1\leq i <k$ there is a $(p+1)$-cell $\tau$ which contains $\sigma_i$ and $\sigma_{i+1}$ satisfying $\{\sigma_i,\tau\}\in \M$ and $\{\sigma_{i+1},\tau\}\in H(\K)$. If $\sigma_1=\sigma_k$, the $\M$-path is \textit{closed}. A discrete \textit{acyclic} vector field is a discrete vector field containing no closed~$\M$-path.

\begin{figure}[!ht]
\begin{center}
\begin{tabular}{cc}
\includegraphics[width=5.9cm]{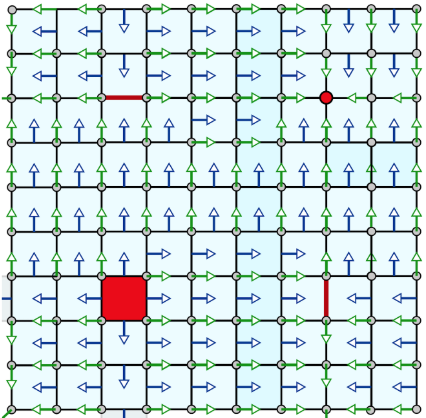}&
\includegraphics[width=5.9cm]{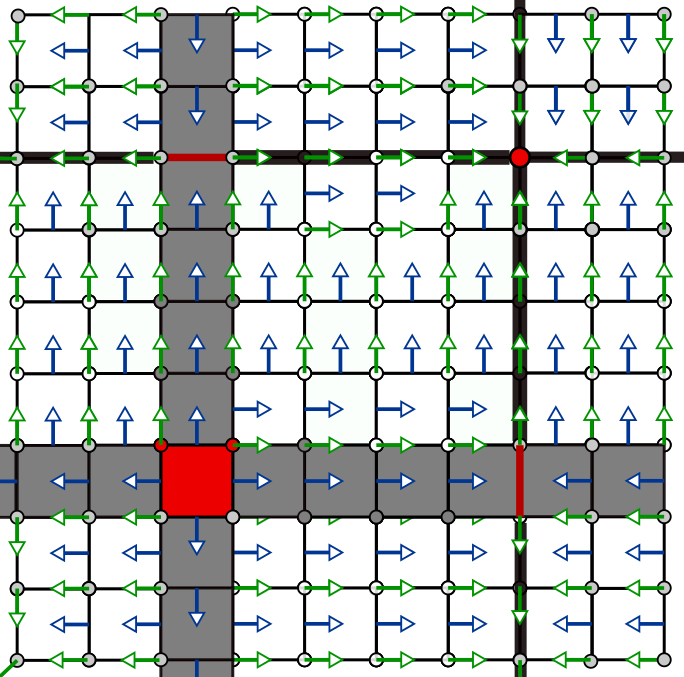}\\
\,(a)&(b)
\end{tabular}
\caption{(a) A discrete vector field. (b) Its topological graph.}
\label{figure:DVF_DECOMPOSITION}
\end{center}
\end{figure}

The critical elements and some special $\M$-paths in a discrete acyclic vector field $(\K,\M)$  provide a \textit{topological graph} $M=(\overline{V},\overline{E})$, a frequent construction in the literature of topological data analysis \cite{tricoche2002}. More precisely, the vertices $\overline{V}$ are the critical elements of $(\K,\M)$. 
An edge in $\overline{E}$ connecting a critical $(p+1)$-cell $\tau$ to a critical $p$-cell $\sigma$ indicates that there exists a $\M${-path} between a $p$-cell of the boundary of $\tau$ and $\sigma$~(Fig.~\ref{figure:DVF_DECOMPOSITION}).

The topological graph splits the original field $(\K,\M)$ into components bounded by $\M${-paths}, giving rise to the {\it Morse--Smale decomposition} of $(\K,\M)$.

\begin{theorem}[Morse--Smale decomposition]
	The critical elements and the $\M$-paths produce a Morse--Smale decomposition of the discrete vector field in regions containing no critical cells. 
\end{theorem}

The Morse--Smale decomposition of a discrete acyclic vector field $(\K,\M)$ can be computed in linear time \cite{lewiner2013critical}, leading to their use in geometry processing applications in surfaces \cite{lewiner2005,lewiner2013critical,lewiner2003,jan11,jan11_1,weinkauf2010topology} and $3$-manifolds \cite{weinkauf2005}. 

The above result is a consequence of Forman's homotopy theorem, a discrete version of the basic homotopy theorem of Morse theory~\cite{milnor16}.  Forman's theorem states that every (regular) CW complex with an acyclic discrete vector field is homotopy equivalent to a CW complex with one $d$-cell for each critical $d$-cell. This result is equivalent to the following statement.

\begin{theorem}[Homotopy]
\label{theorem:forman_homotopy}
Given a discrete acyclic vector field $(\K,\M)$ of $S$ there is another discrete acyclic vector field $(\overline{\K}, \emptyset)$  of $S$ with the same topological~graph. 
\end{theorem}	
In other words, $(\K,\M)$ and $(\overline{\K}, \emptyset)$ have the same critical cells (as in the original statement of Homotopy theorem) and $\M$-paths.

The cancellation of critical elements in a discrete vector field is another tool in Forman's theory with a number of applications, e.g. in the topology simplification of vector fields \cite{weinkauf2010topology}, noise removal \cite{jan11,weinkauf2010topology, weinkauf2005}, and persistent homology \cite{mischaikow2013morse}. To describe it, we use an operation in $\M$-paths connecting critical cells.

\begin{figure}[!ht]
\begin{center}
\begin{tabular}{cc}
\includegraphics[width=5.cm]{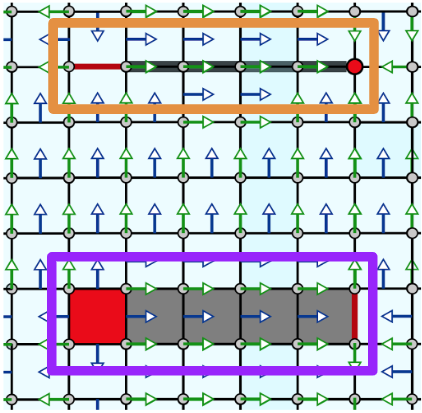}&
\includegraphics[width=5.cm]{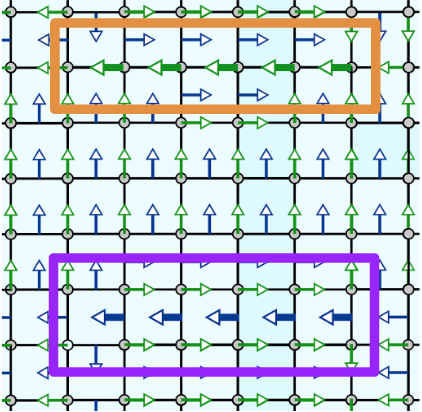}\\
\,(a)&(b)
\end{tabular}
\caption{(a) Two $\M$-paths of dimension $0$ and $1$,  connecting critical elements (in red).\newline (b)~Reversal of both $\M$-paths reduces the number of critical elements.}
\label{figure:DVF_cancellation}
\end{center}
\end{figure}

Consider a $\M$-path $\gamma=\sigma_1\sigma_2\sigma_3\ldots\sigma_k$ connecting a critical $(p+1)$-cell $\tau$ and a critical $p$-cell $\sigma$, where $\tau$ contains $\sigma_1$ on its boundary and $\sigma_k=\sigma$ (Fig.~\ref{figure:DVF_cancellation}(a)). We \textit{reverse} $\M$-path $\gamma$ as follows. By definition, for each $1\leq i <k$ there is a $(p+1)$-cell $\omega$ which contains $\sigma_i$ and $\sigma_{i+1}$ satisfying $\{\sigma_i,\omega\}\in \M$ and $\{\sigma_{i+1},\omega\}\in H(\K)$. To reverse $\gamma$, remove $\{\sigma_i,\omega\}$ from $\M$ and  add $\{\sigma_{i+1},\omega\}$ and the matching $\{\sigma_1,\tau\}$ to $\M$ (Fig.~\ref{figure:DVF_cancellation}(b)). Forman \cite{forman98} proved that when $\gamma$ is the unique $\M$-path between $\tau$ and $\sigma$, reversing $\gamma$ does not create a closed $\M$-path.

\begin{theorem}[Cancellation]
\label{theorem:forman_cancellation}
Two critical cells connected by a unique $\M$-path can be cancelled in a discrete vector field.
\end{theorem}

\vspace{0.5cm}
\section{Discrete line fields}\label{discretelinefields}\hfill\\
The definition of a discrete line field requires an object which generalizes the concept of a discrete vector field with no references whatsoever to directionality.

\begin{figure}[!ht]
	\begin{center}
		\includegraphics[width=1.8cm]{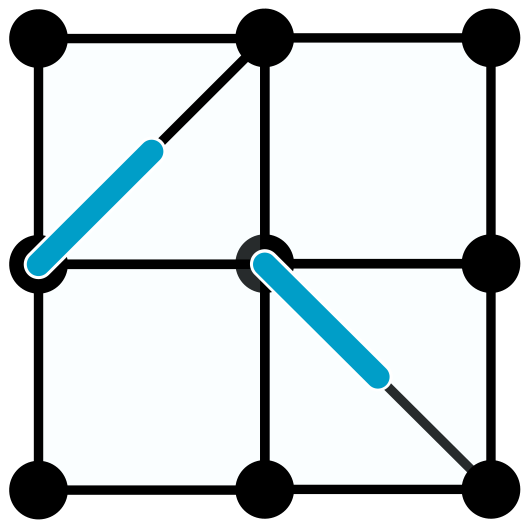}
		\vspace{-0.2cm}
		\caption{The blue lines joining adjacent 	vertices and edges represent the matching.}
		\label{toro_dlf}
	\end{center}
\end{figure}

Let $\K=(V,E,F)$ be a CW decomposition of a compact surface $S$. A {\it discrete line field} is a pair $(\K,\L)$ where $\L$ consists of a matching between vertices and edges of $\K$. As a simple example, consider the  discrete line field on the torus in Fig. \ref{toro_dlf}: the square is interpreted with the usual identifications on the boundary. Notice that not all vertices and not all edges are matched.

In order to convince the reader that such a simple definition is mathematically rich (and appropriate to our purposes), we start showing how Forman's discrete vector fields give rise to line fields in an injective fashion up to duality. In particular, constructions in the literature for discrete vector fields \cite{jan11,robins2011theory,lewiner2003,adiprasito2013metric,gyulassy2008combinatorial, paixao14,paixao2018greedy,  reininghaus2012computational} may be pushed to this subclass of line fields.

\subsection{Discrete vector fields as discrete line fields}\label{subsection:vector_to_line}\hfill\\
Recall Pisanski's \textit{radial decomposition} $R(\K) = (V_R,E_R,F_R)$ of a CW decomposition $\K=(V,E,F)$ \cite{pisanski1984}. The set of vertices $V_R$ is the union of $V$ and a point in the interior of each face of $\K$: we abuse the language and denote $V_R=V\sqcup F$. The edges $E_R$ indicate the adjacency relations between vertices and faces in $\K$, and  are represented by disjoint arcs in $S$. The faces $F_R$ are the $2$-cells $S - (V_R,E_R)$, which correspond to the edges of $\K$, since each edge is either a loop or has two vertices and either belongs to a single face or two. 

The graph $(V_R,E_R)$ is bipartite and all the faces $F_R$ are  quadrilaterals by definition. According to Pisanski \cite{pisanski1984}, a CW decomposition $R = (V_R,E_R,F_R)$ endowed with these properties is the radial decomposition of some CW decomposition $\K$. The radial decomposition $R(\K)$  encodes both CW decompositions $\K$ and its dual $\K^*$, since by construction their radial decompositions coincide.
Thus, discrete vector fields are mapped up to duality (the discrete counterpart of no directionality) to a restricted class of discrete line fields.

\begin{figure}[!ht]
\begin{center}
\begin{tabular}{cc}
\includegraphics[width=5.1cm]{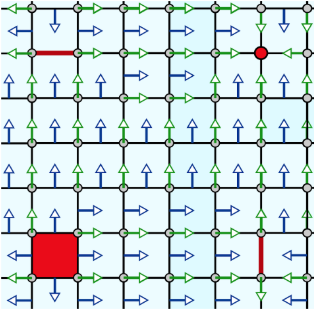}&
\includegraphics[width=5.cm]{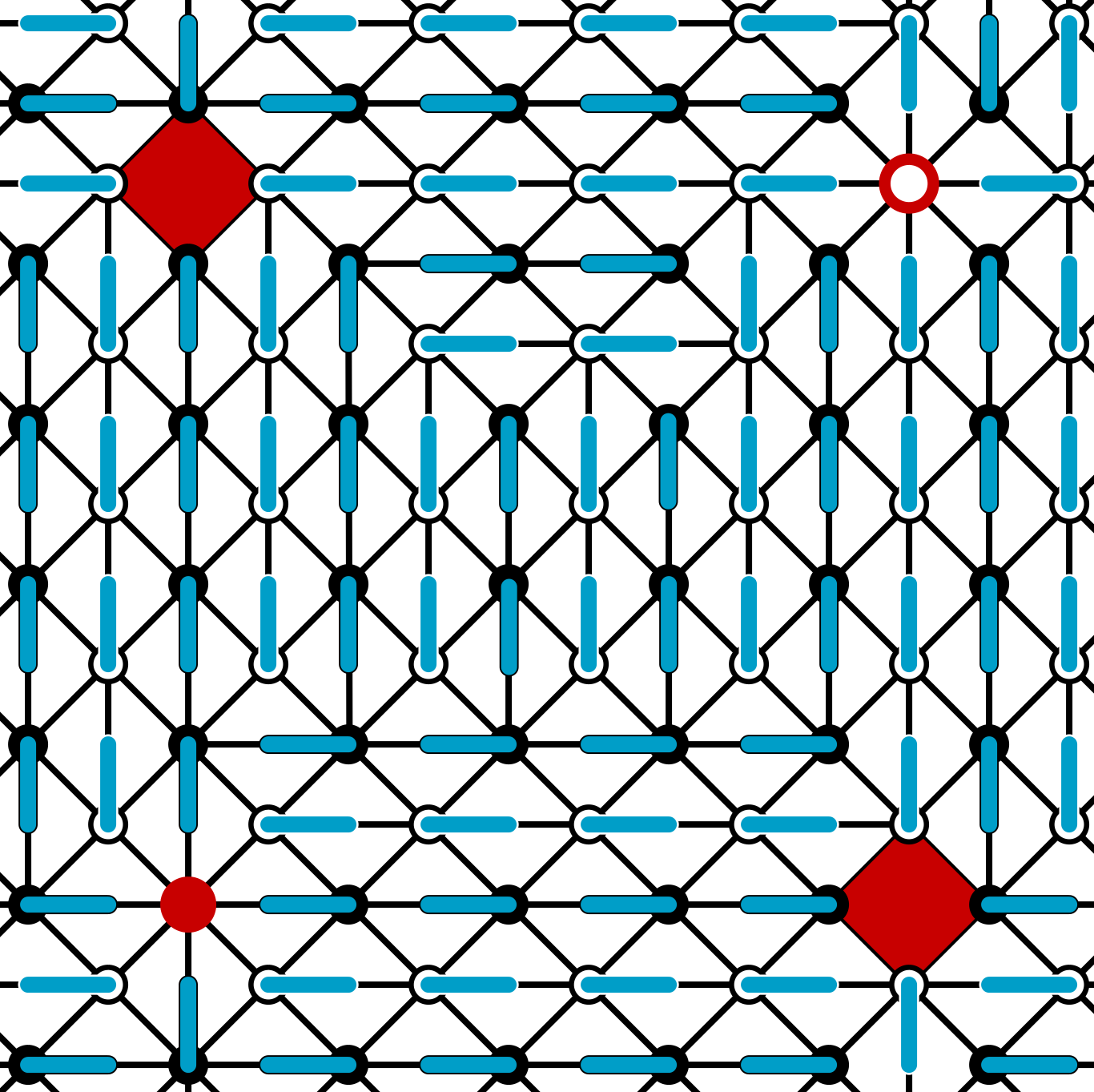}\\
\,(a)&(b)
\end{tabular}\vspace{-0.1cm}
\caption{(a) A discrete vector field. (b) Its correspondent discrete line field.}
\label{figure:DVF_DLF}
\end{center}
\end{figure} 
\vspace{-0.1cm}
\begin{theorem}
\label{theorem:DVF_DLF}
There is an injective map taking  pairs of discrete vector fields $((\K,\M),(\K^*,\M^*))$ to discrete line fields $(R,\L)$. The unmatched edges in $(R,\L)$ define the radial decomposition of $\K$ and $\L$ in turn obtains $\M$.  
\end{theorem}

The identification of a discrete vector field with its dual under the map above is in accordance with the related property in the continuous case. We consider two discrete vector fields $(\K_1,\M_1)$ and $(\K_2,\M_2)$ of a surface $S$ equal if there is an homeomorphism of $S$ taking $\K_1$ to $\K_2$ and $\M_1$ to $\M_2$. Equality for discrete line fields is analogous.

\begin{figure}[ht]
	\begin{center}
			\begin{tabular}{ccc}
				\includegraphics[width=1.4cm]{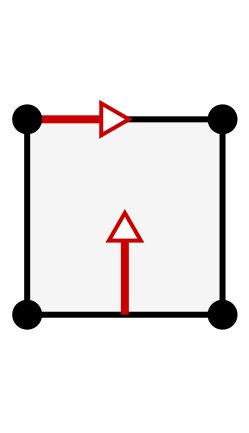}&
				\includegraphics[width=2.4cm]{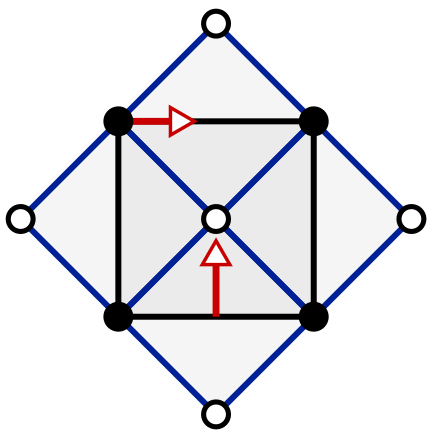}&
				\includegraphics[width=2.4cm]{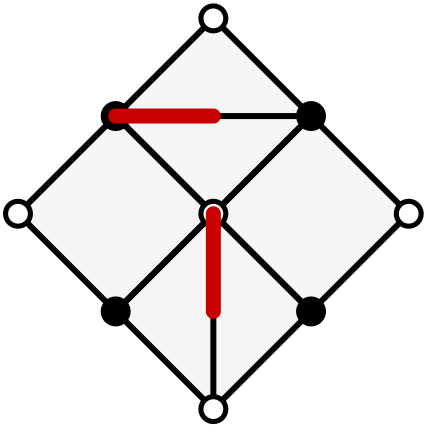}\\
				(a)&(b)&(c)
		\end{tabular}
		\caption{(a) A discrete vector field. (b) The overlap of (a) and its radial graph. (c) The corresponding discrete line field.}
		\label{figure:MMatching_DVF}
	\end{center}
\end{figure}

\begin{proof}
We construct the discrete line field $(R,\L)$ from the radial decomposition $R(\K)$ of $\K$.
Let $\{\sigma,\tau\}\in \M$ be a matching of an edge $\sigma$ and a face (vertex) $\tau$ (Fig. \ref{figure:MMatching_DVF}(a)). In $R(\K)$, $\sigma$ and $\tau$ correspond to a face $f$ and a vertex $w$ (Fig. \ref{figure:MMatching_DVF}(b)). Match  $w$ and its adjacent diagonal $e$ of $f$ (Fig. \ref{figure:MMatching_DVF}(c)) creating $\{w,e\}$. Repeat this construction for each matching in $\M$, resulting in an acyclic matching $\L$ between vertices and~face diagonals in $R(\K)$. Add these face diagonals to the decomposition $R(\K)$ to obtain~a decomposition $R$ which, together with $\L$, gives rise to the desired discrete line field~$(R,\L)$. 

By construction, the unmatched edges of $(R,\L)$ define the radial decomposition of $\K$ which, as shown by Pisanski \cite{pisanski1984}, is in bijection with both $\K$ and its dual $\K^*$, so that $\L$ in turn obtains $\M$. Injectivity then follows.
\end{proof}

We make a remark about identifying critical cells between both fields.
The critical vertices and faces of $(\K,\M)$ correspond to the unmatched vertices in $(R,\L)$ (given by Theorem \ref{theorem:DVF_DLF}), since faces and vertices of $\K$ correspond only to vertices of $R(\K)$. Moreover, the critical edges in $(\K,\M)$ join the quadrilaterals in $(R,\L)$. In Fig. \ref{figure:DVF_DLF}, such cell correspondences are marked in red.

The map in Theorem \ref{theorem:DVF_DLF} is clearly not surjective: some discrete line fields are not discrete vector fields. Simply match any edge of the discrete field in Fig.~\ref{figure:MMatching_DVF}(c).

\subsection{Euler--Poincar\'e formula for discrete line fields}\hfill\\
A vertex $v$ in a discrete line field $ (\K, \L) $ is {\it critical} if it is unmatched in $\L$. The {\it index} of a vertex $v$ is $1$ if it is critical and $0$ otherwise. 
Let $c(f)$ be the number of unmatched edges in the boundary walk of a face $f$. We say that $f$ is {\it critical} if $c(f)\neq 2$. Its index is $1-\frac{c(f)}{2}$. 
There are no critical edges. The critical cells in the discrete line field in Fig. \ref{figure:DVF_DLF} are marked in red.
 
\begin{theorem}
\label{theorem:euler_formula}
Let $\K=(V,E,F)$ be a decomposition of the compact surface $S$ and $(\K,\L)$ be a discrete line field. Then
$$\displaystyle\chi(S)= \sum_{v\in V}index(v)+\sum_{f\in F}index(f).$$
\end{theorem}
The proof will be given after  Theorem \ref{theorem:homotopy}.

\subsection{The Morse--Smale decomposition}
\begin{figure}[!ht]
	\begin{center}
		\begin{tabular}{cc}
			\includegraphics[width=6.cm]{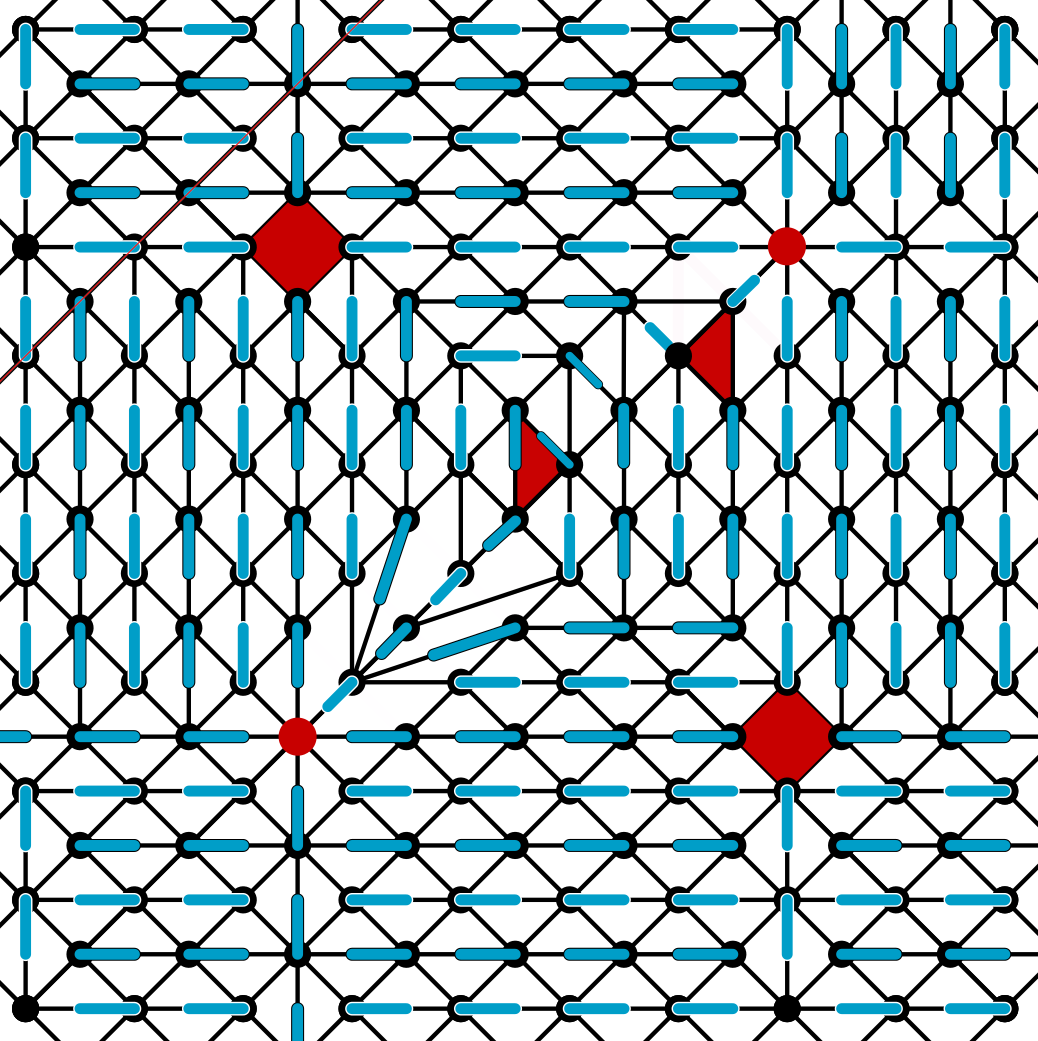}&
			\includegraphics[width=6.cm]{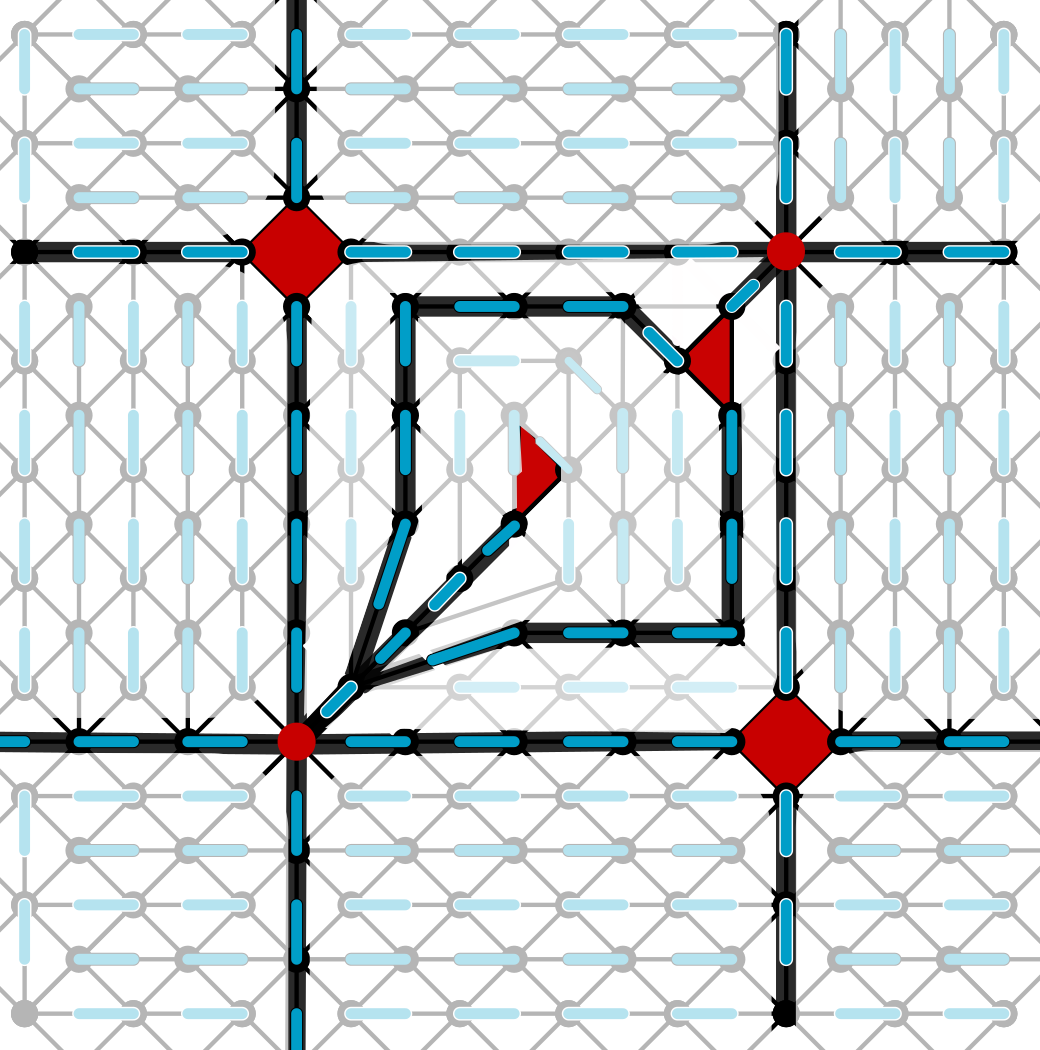}\\[-0.ex]
			(a)&(b)
		\end{tabular}
		\caption{(a) A discrete line field,  critical cells  in red. (b) The Morse--Smale decomposition.}
		\label{figure:sphere_complex_linefield_Ggraph_morsesmale}
	\end{center}
\end{figure}

\hfill\\
Fig. \ref{figure:sphere_complex_linefield_Ggraph_morsesmale}(a) provides an example of a discrete line field: the blue lines represent the matching between vertices and edges in the decomposition.

Again, the basic ingredient in the construction of a \textit{Morse--Smale decomposition}  is the definition of an \textit{$\L$-path},  a sequence  of vertices in $\K$,
$\gamma=v_1v_2v_3\ldots v_k,$
such that for each $1\leq i <k$ there is a edge $e$ incident to $v_i$ and $v_{i+1}$ satisfying $\{v_i,e\}\in \L$. If $v_1=v_k$, the $\L$-path is \textit{closed}. A discrete line field is \textit{acyclic} if it has no closed $\L$-path.

Morever, the \textit{topological graph} $M=(\overline{V},\overline{E})$ of a discrete acyclic line field $(\K,\L)$ is defined by the critical elements, which form $\overline{V}$, and some special $\L$-paths   (Fig.~\ref{figure:sphere_complex_linefield_Ggraph_morsesmale}(b)).
An edge in $\overline{E}$ connecting a critical face $f$ to a critical vertex $v$ indicates that there exists a $\L${-path}, containing no edge in the boundary walk of $f$, between a vertex~of~$f$~and~$v$.


\begin{theorem}[Morse--Smale decomposition]
\label{theorem:MS_decomposition_line}
	The critical elements and the $\L$-paths produce a $($Morse-Smale$)$ decomposition of the discrete line field in regions containing no critical cells. 
\end{theorem}

We postpone the proof until we state Theorem \ref{theorem:homotopy}. Again, the Morse--Smale decomposition of a discrete line field can be computed in linear time because $\L${-paths} are obtained in a similar manner to the case of discrete vector fields~\cite{lewiner2013critical}.

\subsection{A homotopy theorem and the missing proofs}
\begin{theorem}
\label{theorem:homotopy}
Given a discrete acyclic line field $(\K,\L)$ of $S$, there is another discrete acyclic line field $(\overline{\K},\emptyset)$ of $S$ with the same topological graph.
\end{theorem}
\begin{proof}
Let $\{\sigma, \tau\}$ be a matched pair in $\L$, for a vertex $\sigma$ and an edge $\tau$. Contracting $\tau$ into a point, we collapse $\sigma$ over $\tau$. The result is a new discrete line field with the same topological graph and one less element in the matching. The acyclicity of $\L$ guarantees we repeat this procedure until there are no more matched edges in $\L$.
A non-critical face $f$ has only two different edges $e_1$ and $e_2$ along its boundary walk, since all matched edges have been annihilated, we collapse $e_1$ over $f$. Repeat until there are no more non-critical faces.
\end{proof}

\begin{figure}[!ht]
\begin{center}
\setlength{\tabcolsep}{0.1em}
\begin{tabular}{cccc}
\includegraphics[width=3.1cm]{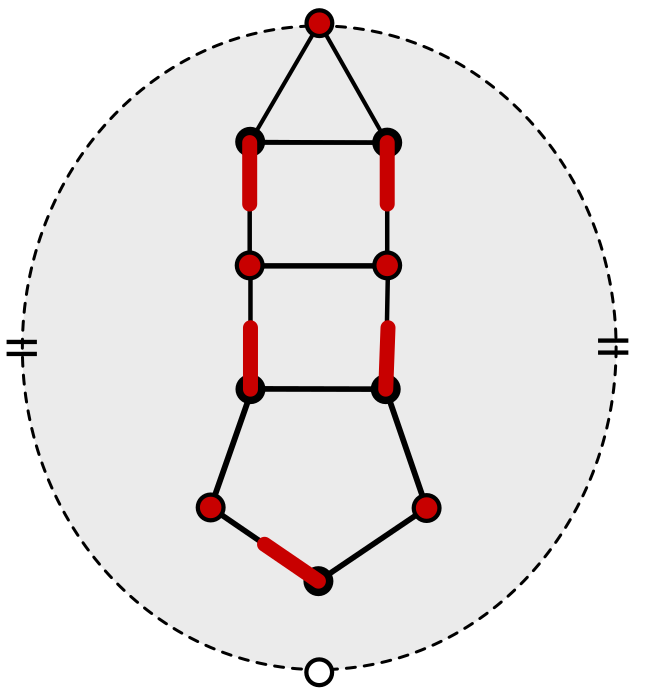}&\includegraphics[width=3.1cm]{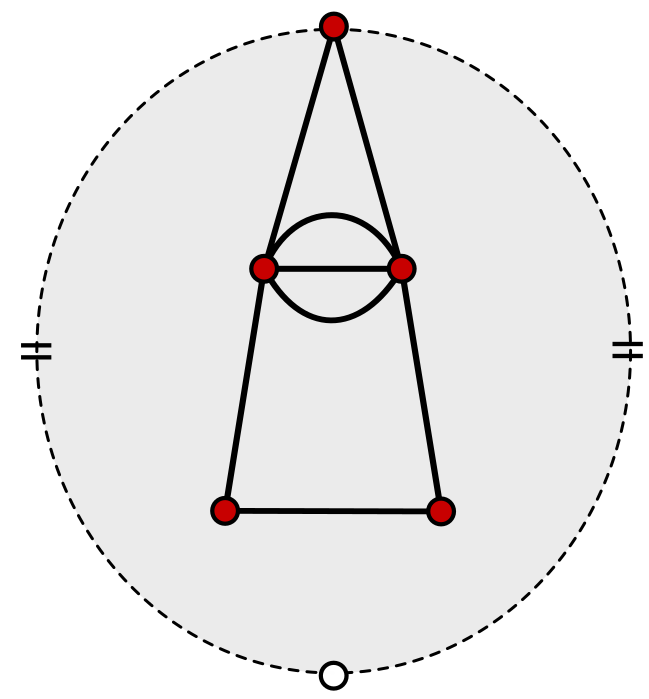}&\includegraphics[width=3.1cm]{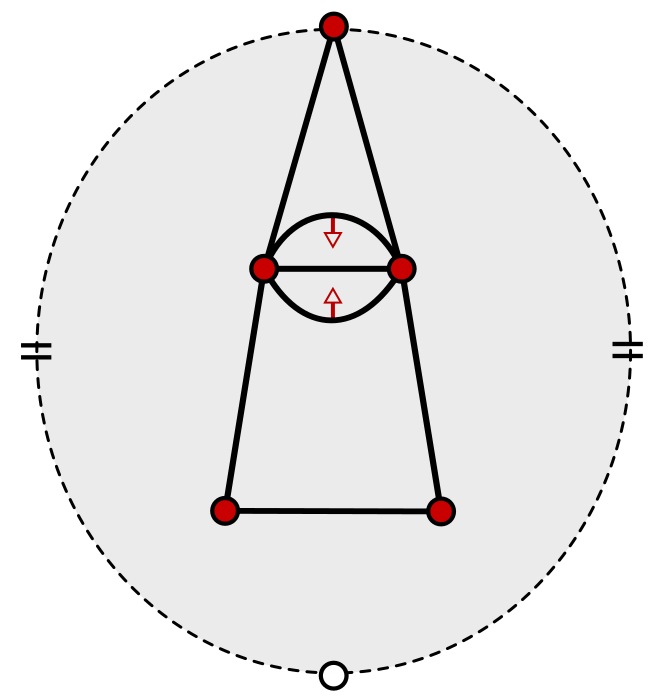}&
\includegraphics[width=3.1cm]{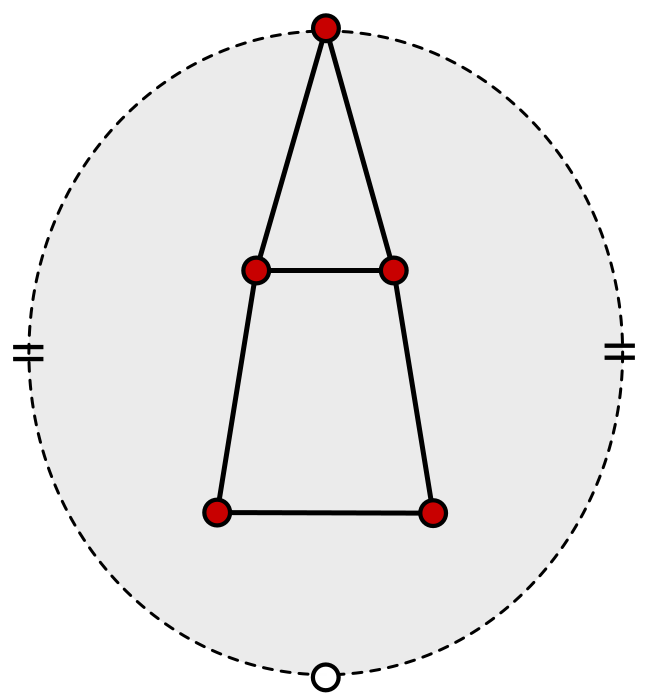}\\
(a)&(b)&(c)&(d)
\end{tabular}
\caption{(a) 
a discrete line field. (b)-(d) Getting rid of non-critical faces.}
\label{figure:last_theorem}
\end{center}
\end{figure} 

From Theorem \ref{theorem:homotopy}, the Morse--Smale decomposition of a discrete acyclic line field $(\K,\L)$ corresponds to the radial decomposition $R(\overline{\K})$ of $\overline{\K}$, since in the discrete line field $(\overline{\K},\emptyset)$ the $\L$-paths indicate the adjacencies between vertices and faces of $\overline{\K}$.

We now prove the Euler--Poincar\'e formula (Theorem~\ref{theorem:euler_formula}) and the Morse--Smale decomposition of a discrete line field (Theorem~\ref{theorem:MS_decomposition_line}).

\bigskip
\begin{proof}[ Proof of Theorem~\ref{theorem:euler_formula}]
The collapses in the proof of Theorem~\ref{theorem:homotopy} preserve the sum $\sum_{}index(v)+\sum_{}index(f)$. This is because only non-critical vertices and faces are annihilated during
the procedure. Also, the indices of the critical faces are preserved since only matched edges are collapsed along they boundary walk.
Thus from Theorem \ref{theorem:homotopy}, it is enough to prove the case $\L=\emptyset$. By  Euler's formula for the decomposition $\K=(V,E,F)$,
$$\chi(\K)=|V|-|E|+|F|.$$
Since all vertices are critical, $|V|=\displaystyle \sum_{v\in V}index(v)$. In the sum
$$
\displaystyle \sum_{f\in F}index(f) = |F|-\sum_{f\in F} \frac{c(f)}{2},
$$ 
each edge $e\in E$ appears twice in the boundary walk of the $\K$ face's, since $\K$ is a decomposition of a compact surface. Then  $\displaystyle \sum_{f\in F} \frac{c(f)}{2}=|E|.$
\end{proof}	

\bigskip

\begin{proof}[ Proof of Theorem~\ref{theorem:MS_decomposition_line}]
We check the topological graph $M$ of the discrete acyclic line field $(\K,\L)$ decomposes the CW decomposition $\K$ in regions having no critical cell, and these regions are bounded by four separatrices. 
Theorem \ref{theorem:homotopy} applied to $(\K,\L)$ produces a discrete line field $(\overline{\K},\emptyset)$ with the same topological graph
$M$: it is the graph of adjacencies between vertices and faces of $\overline{\K}$, since it is an empty matching. This graph in turn coincides with the graph of the radial decomposition $R(\overline{\K})$ of $\overline{\K}$: the desired properties follow.
\end{proof}

\subsection{Cancellation in discrete line fields}\label{chapter:cancellation} \hfill\\
We propose two forms of cancellation of critical elements in a discrete line field: a merge between two critical faces which belong to the boundary of a unique region of the Morse--Smale decomposition, and a cancellation between a critical face and a critical vertex connected by a unique path. 

\begin{figure}[!ht]
	\begin{center}
		\begin{tabular}{cc}
			\includegraphics[width=4.cm]{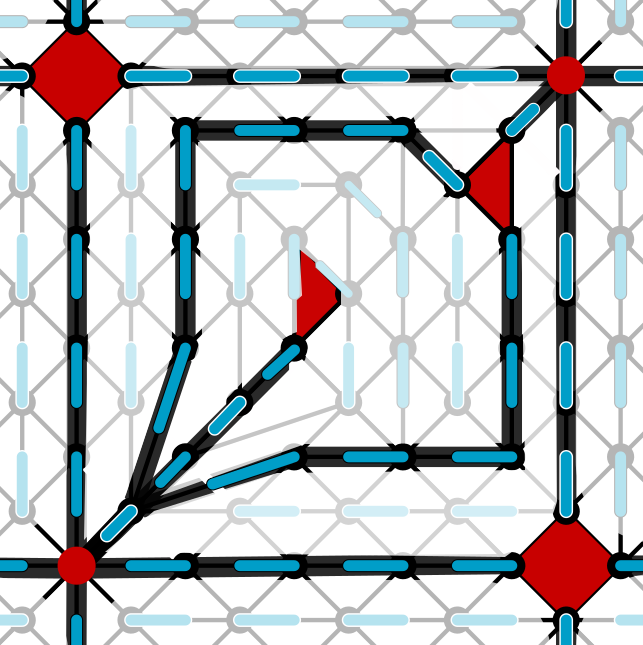}&
			\includegraphics[width=4.cm]{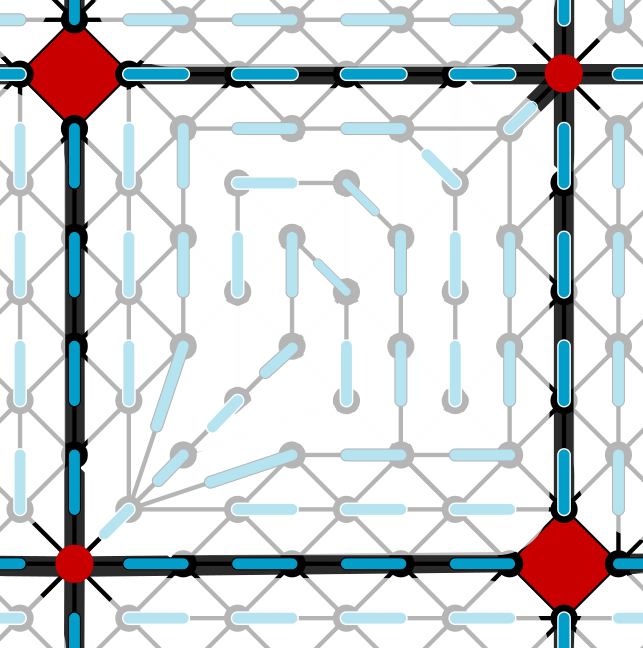}\\
			(a)&(b)
		\end{tabular}
		\caption{(a) the two critical triangles in red are in the boundary of a unique face of the Morse--Smale decomposition. (b) the cancellation between the critical faces.}
		\label{figure:linefield_mergingcells}
	\end{center}
\end{figure}

\begin{theorem}\label{proposition:cancellation_two_faces}
Let $f$ and $g$ be critical faces of a discrete acyclic line field $(\K,\L)$, which belong to a unique face $\Lambda$ of the Morse--Smale decomposition. Then removal of the unmatched edges inside $\Lambda$ gives rise to a new discrete acyclic line field in which $f$ and $g$ disappear: the number of critical cells decreases.
\end{theorem}
\begin{proof}
Being opposite vertices of a unique face $\Lambda$, $f$ and $g$ share a unique edge $a$ in $\overline{\K}$ (Fig.~\ref{figure:linefield_mergingcells}(a)). Thus there are two edges $e\in f$ and $d\in g$ and two sequences of matchings
$\{e,f_1\}, \{e_1, f_2\},\dots,\{e_{k-1}, f_k\}$ and $\{d, g_1\}, \{d_1, g_2\},\dots,\{d_{l-1}, g_l\},$
where both $f_k$ and $g_l$ contain the edge $a$. 
From the uniqueness of the adjacency between $a$ and $f$ in $\overline{\K}$, reversal of the sequence of acyclic matching $\{e, f_1\}, \{e_1, f_2\},\dots,\{e_{k-1}, f_k\}$ does not create closed paths in the acyclic matching. For the reversed sequence $\{a, f_k\}, \{e_{k-1}, f_{k-1}\},\dots,\{e_{1}, f_1\}, \{e, f\}$, $$\left\{\{d, g_1\}, \{d_1, g_2\},\dots,\{d_{l-1}, g_l\},\{a, f_k\}, \{e_{k-1}, f_{k-1}\},\dots,\{e_{1}, f_1\}, \{e, f\}\right\}$$ is an acyclic matching on $H(\K)$: Theorem \ref{theorem:forman_homotopy} completes the proof  (Fig.~\ref{figure:linefield_mergingcells}(b)).
\end{proof}

\begin{figure}[!ht]
	\begin{center}
		\begin{tabular}{cc}
			\includegraphics[width=5.cm]{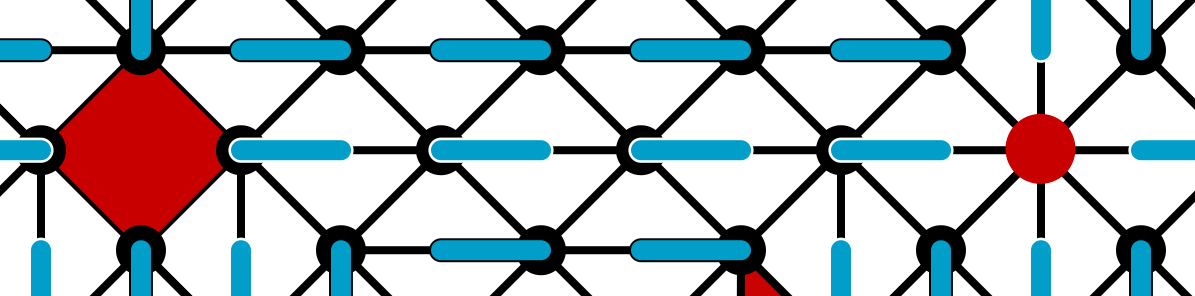}&
			\includegraphics[width=5.cm]{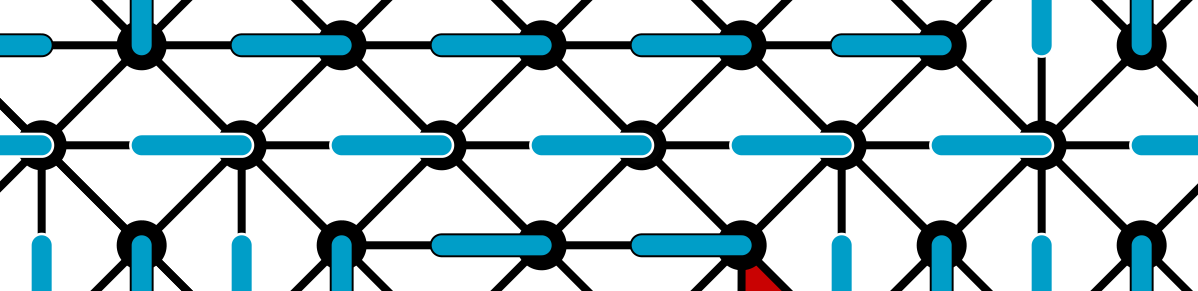}\\
			(a)&(b)
		\end{tabular}
		\caption{ (a) Critical face and vertex connected by a unique  path. (b) The reversed  path.}
		\label{figure:linefield_cancelling}
	\end{center}
\end{figure}

The cancellation between a vertex $v$ and a face $f$ in a discrete line field $(\K,\L)$ is based on Forman's cancellation of critical elements in a discrete acyclic vector field (Theorem \ref{theorem:forman_cancellation}). A  path $v=v_{1}v_{2}\dots v_{k}$ from $v_1$ to a vertex $v_k$ in the boundary walk of $f$ can be reversed as follows. For each $1\leq i<k$ there is an edge $e_{i}$ which runs from $v_{i}$ to $v_{i+1}$, satisfying $\{v_{i},e_{i}\}\in \L$. To reverse the path, for $i=2,\ldots,k-1$, remove  $\{v_{i},e_{i}\}$ from $\L$ and add $\{v_{i+1},e_{i}\}$ to $\L$, also add $\{v_1,e\}$ to $\L$, where $e$ is a diagonal of $f$ incident to $v_1$ (Fig. \ref{figure:linefield_cancelling}). 

\begin{theorem}\label{proposition:cancellation_vertex_face}
Let $v$ be a critical vertex and $f$ a critical face with $index(f)<0$,  connected by a unique  $\L$-path $\gamma$.
Reversal of $\gamma$ subdivides $f$ into two faces $f_0\cup e \cup f_1$ and  $v$ and $f_0$ are non-critical cells of the resulting line field.
\end{theorem}
\begin{proof}
As $index(f)<0$, the boundary walk of $f$ has more than $3$ unmatched edges: choose the diagonal $e$ of $f$ such that, in the subdivision $f_0\cup e \cup f_1$, the face $f_0$ has only two unmatched edges on its boundary walk.
Reversal of the path $\gamma$ does not create a cycle, since otherwise there would be another path between $v$ and $f$, contradicting the hypothesis.
\end{proof}

\section{Conclusions}\hfill\\
We introduced discrete line fields, with properties akin to those obtained for Forman's discrete vector fields. We hope that the simplicity and the combinatorial nature of our definition lead to further exploration of theoretical and applied possibilities.

\section*{Acknowledgements}\hfill\\
C. Tomei gratefully acknowledges support from CAPES, CNPq and FAPERJ. T. Novello was
supported by a CAPES grant. 
\newpage
\bibliographystyle{siamplain}
\bibliography{references}
\end{document}


\maketitle

\section{A detailed example}

Here we include some equations and theorem-like environments to show
how these are labeled in a supplement and can be referenced from the
main text.
Consider the following equation:
\begin{equation}
  \label{eq:suppa}
  a^2 + b^2 = c^2.
\end{equation}
You can also reference equations such as \cref{eq:matrices,eq:bb} 
from the main article in this supplement.

\lipsum[100-101]

\begin{theorem}
  An example theorem.
\end{theorem}

\lipsum[102]
 
\begin{lemma}
  An example lemma.
\end{lemma}

\lipsum[103-105]

Here is an example citation: \cite{KoMa14}.

\section[Proof of Thm]{Proof of \cref{thm:bigthm}}
\label{sec:proof}
\lipsum[106-112]

\section{Additional experimental results}
\Cref{tab:foo} shows additional
supporting evidence. 

\begin{table}[htbp]
{\footnotesize
  \caption{Example table}  \label{tab:foo}
\begin{center}
  \begin{tabular}{|c|c|c|} \hline
   Species & \bf Mean & \bf Std.~Dev. \\ \hline
    1 & 3.4 & 1.2 \\
    2 & 5.4 & 0.6 \\ \hline
  \end{tabular}
\end{center}
}
\end{table}

\bibliographystyle{siamplain}
\bibliography{references}